\documentclass[10pt, twoside]{article}

\usepackage[numbers]{natbib}
\usepackage[affil-it]{authblk}

\usepackage[T1]{fontenc} 
\usepackage{microtype} 
\usepackage[columnsep=25pt]{geometry}
\usepackage{multicol} 
\usepackage[hang, small,labelfont=bf,up,textfont=it,up]{caption} 
\usepackage{booktabs} 
\usepackage{float} 

\usepackage{abstract}

\usepackage{titlesec} 
\renewcommand\thesection{\Roman{section}} 
\renewcommand\thesubsection{\Roman{subsection}}
\titleformat{\section}[block]{\large\scshape\centering}{\thesection.}{1em}{}
\titleformat{\subsection}[block]{\large}{\thesubsection.}{1em}{} 

\usepackage{fancyhdr}
\usepackage{amsmath}
\usepackage{amsthm}
\usepackage[english]{babel}
\usepackage{graphicx}
\usepackage{anysize}
\usepackage{amssymb} 
\usepackage{float}
\usepackage{times}
\usepackage{mathtools}
\usepackage{color}
\usepackage{mathrsfs}

\usepackage{hyperref} 

\newcommand{\PD}[2]{\frac{\partial #1}{\partial #2}}

\newcommand{\twovector}[2] {\ensuremath{\left(\begin{array}{c} #1 \\ #2 \end{array}\right)}}

\title{Non-Quantum Behaviors of Configuration-Space Density Formulations of quantum mechanics}
\author{\large Philipp Roser\thanks{roserp@wwu.edu}\, and Matthew T. Scoggins\thanks{scoggim@wwu.edu}}
\affil{\small Department of Physics and Astronomy,\\ Western Washington University\\ Bellingham, WA 98225, USA}
\date{\vspace{-0.5cm}}		

\begin{document}
\maketitle
\begin{abstract}
The trajectories of the pilot-wave formulation of quantum mechanics and hence its empirical predictions may be recovered via the dynamics of a density function on the configuration space of a system, without reference to a physical wave function. We label such formulations `\textit{CSD} frameworks.' But this result only holds if a particular, apparently \textit{ad hoc} condition, broadly speaking equivalent to the single-valuedness of the wave function in standard quantum mechanics, is imposed. Here we relax this condition. We describe the types of scenarios in which this would lead to deviations from quantum mechanics. Using computational models we ask how the degree of `non-quantumness' of a state, suitably defined, changes with time. We find that it remains constant in time even under non-trivial dynamics, and argue that this implies that a \textit{dynamical} justification of the Wallstrom condition is unlikely to be successful. However, we also make certain observations about stationary states in CSD frameworks, which may offer a way forward in justifying the Wallstrom condition.
\end{abstract}

\begin{multicols}{2}

\section{Introduction}\label{sec.introduction}
Quantum mechanics is well established as a framework of extraordinary empirical success. Its conceptual foundations however remain a matter of the debate. The challenge to provide a physical interpretation of the mathematical structure of quantum theory has engendered a great variety of "interpretations", "models" and "approaches" over the last one hundred or so years. The literature on these questions is vast.\footnote{No reasonably sized list of references would do this justice. For an overview of traditional approaches, see e.g.\ \cite{Wallace2008}. A reader looking for a broad exposure to different approaches may also find \cite{Schlosshauer2011} useful.}

In at least one such family of approaches \cite{Tipler2010,Sebens2014,HallDeckertWiseman2014, Schmelzer2011a, Smolin2015,Bostrom2014,Roser2015_TrajGeometry} the wave function is not considered a physically fundamental entity. 
Instead, a continuous\footnote{Some authors (\cite{Sebens2014,HallDeckertWiseman2014} however argue that the uncountably infinite number of systems in configuration space is only an approximation to a dense but countable number of systems for reasons we will not critique here. The resulting theory would be experimentally distinguishable from standard quantum theory, at least in principle.} ensemble of interacting ``classical-like'' systems of varying density in configuration space forms the basic ontological foundation, together with a field of configuration-space velocities. From these the wave function can be reconstructed mathematically (up to one single degree of freedom, the global phase). This ensures compliance with the PBR theorem \cite{PuseyBarrettRudolph2012}, for example. Different authors have referred to these approaches variously by ``Hamilton-Jacobi Many-World Theory'' \cite{Tipler2010},"Many Interacting Worlds'' \cite{HallDeckertWiseman2014}, and other names, although confusion with the conceptually very different many-worlds theory of Everett \cite{Everett1956} needs to be avoided. For this reason we will label such approaches simply as configuration-space density (CSD) frameworks. The origins of this type of foundation can be traced back to (incomplete) ideas of Madelung's ``hydrodynamical'' approach \cite{Madelung1926}. A more detailed summary of the similarities and differences between frameworks within this family is found elsewhere \cite{Roser2015_TrajGeometry}. 

CSD approaches all share one feature: In order to recover the phenomenology of quantum mechanics, a quantization condition must be added ``by hand'', namely the condition that the line integral of the velocity field in configuration space around any closed loop can only take discrete values. This condition is key for the quantized nature of angular-momentum values, for example. In other frameworks this condition is guaranteed by the single-valuedness of the wave function, which is absent in CSD approaches. This necessary \textit{ad hoc} addition detracts from the conceptual appeal of CSD theories.

In this paper we consider the consequences of relaxing this condition in the context of a CSD approach to non-relativistic quantum mechanics. We ask if the ``non-quantumness'' of this generalized framework actually poses an empirical problem, or if it in some way disappears dynamically. We find that at least in the context of fairly simple systems the problem does \textit{not} disappear dynamically and the challenge posed to CSD-type models remains.

In section \ref{sec.QMasQHJT} we summarize the CSD framework and make some general observations. In section \ref{sec.wallstromnonquamtum} we explain the problem of quantized values in situations with periodic boundary conditions. We also identify in what sort of systems the problem arises and highlight some more subtle aspect of the issue. In section \ref{sec.numericalanalysis} we describe how one may approach an investigation of the wider dynamics that CSD allows, and in section \ref{sec.casestudies} we present the specific case study of the 2D harmonic oscillator. Section \ref{sec.stationarystates} constitutes a discussion of stationary states in CSD frameworks, which may yet provide insights with regards to the condition. We conclude in section \ref{sec.conclusion}.


\section{Quantum mechanics as a quantum-Hamilton-Jacobi theory of an ensemble}\label{sec.QMasQHJT}

A framework that reproduces the empirical results of (non-relativistic) quantum mechanics may be formulated as follows. Consider a classical $n$-dimensional configuration space that represents the degrees of freedom of the system. For concreteness, consider a system of $N$ particles of mass $m_i$, $i\in\{1, ..., N\}$ in three-dimensional space, $n=3N$, although much of our discussion generalizes freely to other systems. The Schr\"odinger equation for the wave function $\psi$ of this system may be split into its real and imaginary parts, giving the quantum Hamilton-Jacobi equation and a continuity equation for the system:
\begin{align}
       &~&   -\PD{S}{t} = \sum_{I=1}^N\frac{(\nabla_I S)^2}{2m_I} + V(x) + Q\big[|\psi|\big] \\
       &~& \PD{|\psi|^2}{t}+\nabla\cdot(|\psi|^2\nabla_I S/m_I) = 0 \label{CE}
\end{align}
where $Q = -\sum_{I=1}^N\frac{\hbar^2}{2m_I}(\nabla_I^2|\psi|)/|\psi|$ is the quantum potential. Here $V$ is the potential function describing this system and $S$ is the phase of $\psi$. This constitutes the entry point to Bohm's second-order formulation of pilot-wave theory \cite{Bohm1952a,Bohm1952b,Holland1993}. If a large number of identical systems described by these equations form a density $\rho$ in configuration space, one can readily show that \emph{if} at some initial time $t_0$ it is the case that $\rho=|\psi|^2$ and that the configuration-space velocity of the system is $\vec{v}_i=\nabla_iS/m_i$, \emph{then} the equality $\rho=|\psi|^2$ will continue to hold at all times, and the empirical predictions of quantum mechanics can be recovered. There are, of course, various wider conceptual issues concerning the interpretation of probability that may be relevant to a more thorough discussion here, but we leave these aside in the present paper and instead refer the reader to the extensive literature on the subject.

The question arises how to explain the equality $\rho=|\psi|^2$ for some initial time. Some authors have proposed typicality arguments \cite{DuerrGoldsteinZanghi1992}. However, it has also been shown theoretically \cite{Valentini1991a} and through extensive simulation \cite{ValentiniWestman2004} that the condition $\rho=|\psi|^2$ constitutes a thermodynamic-like equilibrium: If $\rho\neq|\psi|^2$ at some early time, then under suitably general conditions $\rho$ \emph{approaches} $|\psi|^2$. It may therefore be suggested that if $\rho$ (for some type of system, or the universe) was in this quantum \emph{non}-equilibrium at some point in the distant past, $\rho$ today would be in practice experimentally indistinguishable from exact equality with $|\psi|^2$  \cite{Valentini1991a, ValentiniWestman2004}. It should be noted that the velocity condition $\vec{v}_I=\nabla_IS/m_i$ on the other hand is unstable and deviations from it at some initial time will lead to chaotic behaviour \cite{ColinValentini2014}.

Here we consider a different proposal. We can eliminate any reference to $\psi$ entirely by redefining the quantum potential term as $Q = -\sum_{i=I}^N\frac{\hbar^2}{2m_I}(\nabla^2\sqrt{\rho})/\sqrt{\rho}$ and re-interpreting $S$ as a Hamilton principal function analogue for this system. This constitutes the dynamics in conjunction with the continuity equation for $\rho$. Up to a subtlety whose discussion will constitute much of the present paper, the evolution of the system will remain unchanged from the (equilibrium) pilot-wave prediction. The fundamental dynamical quantities in this framework are the configuration-space density $\rho$ and the velocity field $\vec{v}_i$ (also defined on configuration space and itself a $3N$-dimensional vector field). A quantity $\psi\equiv\sqrt{\rho}e^{iS}$ can be reconstructed mathematically up to a single global phase factor from knowledge of $\rho$ and the spatial gradient of $S$ in the form of the velocity field. The global phase factor is physically irrelevant even in quantum mechanics, so its disappearance poses no issue. 

Formulations along these lines with varying details have been proposed by a number of authors \cite{Tipler2010,Sebens2014,HallDeckertWiseman2014,Schmelzer2011a,Bostrom2014,Roser2015_TrajGeometry,Poirier2010}, and have been employed for their convenience in numerical methods (see \cite{Wyatt2005} for a comprehensive overview). The conceptual foundation can arguably be traced back to Madelung's (incomplete, three-space based) ``hydrodynamic'' approach.

The absence of the wave function does not imply that this theory is ``classical,'' except in the most misleading sense. The dependence of $Q$ on the configuration-space ensemble density $\rho$ has no analogue in classical mechanics. The local configuration-space ensemble density affects motion of a single system in this ensemble. This implies that the other elements of the ensemble must themselves be real constituents of nature, not merely mathematical artifacts. Each point in the configuration space is a ``universe'' or ``world'' that is equal in ontological status to ours. As a result, some \cite{HallDeckertWiseman2014} have dubbed this formulation "many interacting worlds." Care should be taken to distinguish it from the conceptually very different Everettian many-worlds approach \cite{Everett1956}. We will use the term \textit{actual world} in order to distinguish the configuration-space position and trajectory we ``inhabit.''

Aside: ``Ensemble'' in this context does not mean the same as the term in thermodynamics, for example, where an ``ensemble'' refers to a large number of identical systems \emph{in the actual world}, which are not interacting (except possibly through a classical potential) and are described by mathematically identical configuration spaces, which allows their collection to be represented as a density in a single configuration space. The density itself is however irrelevant for the calculation of a single trajectory, in stark contrast with the dynamics of the formulation of quantum mechanics discussed here.

The existence of the quantum potential $Q$ (as a function of configuration-space density rather than of wave function amplitude) can be motivated starting from classical mechanics by imposing the condition that no two configuration-space trajectories intersect, or similarly that the Hamilton Principal Function for the system is differentiable. This condition itself may be motivated by more fundamental conditions, such as determinism \cite{Tipler2010}. We will not critique this discussion here.

One further remark: Rather than beginning with modification of the Hamilton-Jacobi equation, one can introduce the ``force'' $-\nabla_iQ$ at the level of Newton's Law of Motion. The crucial difference is that by doing this we do not assume that the velocity field in configuration space at some initial time $t_0$ is equal to the gradient of some function $S$. That is, we do not assume the existence of a single Hamiltonian Principal Function for all trajectories. This therefore allows for a much wider range of initial conditions for the set of configuration-space trajectories. The situation here also differs from classical mechanics where non-actual trajectories are not presumed physically real, so $S$ is constructed to determine the evolution of our actual system only. 

The mathematical existence of other trajectories evolving according to the same function $S$ in classical mechanics is, in a sense, an afterthought. Furthermore, in contrast with the present framework, the shape of the other trajectories in classical mechanics has no consequence for the configuration-space motion of the actual system. 


\section{Wallstrom's objection and responses, and the relevance of nodes} \label{sec.wallstromnonquamtum}

In the standard quantum formalism, as well as in the Pilot-Wave formulation, discrete sets of values (``quantum numbers'') arise for a large range of systems. Textbook examples include the discrete energy levels of a potential well and the possible angular momentum values of a hydrogen atom. In the first of these two cases the discreteness arises because of boundary conditions imposed on the wave function at the edges of the well, while in the second example, crucially, the \textit{single-valuedness of the wave function} forces the phase to undergo a change that is an integer multiple of $2\pi$ around any closed path (e.g.\ \cite{Holland1993}, although the fact was first noted by Dirac \cite{Dirac1931}),
\begin{equation} \oint_{\text{any loop}}\nabla S\cdot\vec{d\ell} = n\cdot2\pi\hbar,\qquad n\in\mathbb{Z}.\label{eq.WallstromCondition}\end{equation}
Note that the weaker condition that $\nabla S$ be single-valued does not imply that $n$ be an integer. Therefore, if the configuration-space density $\rho$ together with the velocity field $\vec{v}$ are taken to be the fundamental variables, imposing the above condition is unmotivated, except \textit{a posteriori} in order to reproduce (albeit not explain) experimental observation. 

That this requirement needs to be added ``by hand'' in the configuration-space density based approach was first pointed out, to our knowledge, by Wallstrom \cite{Wallstrom1994} as a criticism of Madelung's hydrodynamic formulation. We will therefore refer to condition (\ref{eq.WallstromCondition}) in this context as the \textit{Wallstrom condition}.

Starting in section \ref{sec.numericalanalysis} we will consider the consequences of \textit{not} imposing the Wallstrom condition. For the remainder of the present section we will briefly review how some authors have addressed the issue and then make some more detailed observations about the condition.

Some works proposing a formulation based on configuration-space density fail to address the issue entirely. Others add the Wallstrom condition by hand, acknowledging its \textit{ad hoc} nature (e.g.\ \cite{Sebens2014}), perhaps suggesting that the current theory is only an effective theory and that a more fundamental one may exist from which the Wallstrom condition would follow naturally. 

Among explicit attempts to address the issue are the following. Schmelzer \cite{Schmelzer2011a,Schmelzer2011b} proposes a regularity condition for the Laplacian of $\rho$ and shows that this is mathematically equivalent to demanding that the wave function only have simple zeros, so that the loop integral around any single node is $2\pi\hbar$ exactly. This deviates from the predictions of standard quantum mechanics, which is taken to be an approximation, raising a number of other questions. 

Smolin's proposal \cite{Smolin2015} based on a ``principle of maximum variety'' differs significantly from the others in that it concerns the dynamics of similar systems in our \textit{actual world} and would be experimentally distinguishable from standard quantum mechanics, at least in principle, especially for systems with a small number of identical copies in the universe. We will not discuss its merits in the present paper, but only note that it too encounters the need to explain the Wallstrom condition. The work proposes the existence beables with phase-like properties that take the role of momenta in order to do so.

Elsewhere Roser \cite{Roser2015_TrajGeometry} considered an action integral in which the quantum potential arises as a curvature term in a Weyl geometry (on configuration space), based on earlier insights by Santamato \cite{Santamato1984,Santamato1984b}. Here the function $S$ is introduced by adding the most general possible total time derivative to the Lagrangian. Doing so does not change the equations of motion \cite{Goldstein3rdEd}. The time derivative that is added must be $\mathbb{R}^1$. However, $\mathbb{R}^1$ is the tangent space of \textit{two} possible spaces, $\mathbb{R}^1$ itself and the closed space $\mathbb{S}^1$. The most general total time derivative therefore naturally includes a phase-like function (defined on $\mathbb{S}^1$), giving the function $S$ the properties required to satisfy the Wallstrom condition. Once again we omit further discussion of the details here. 

The problem also arises in Nelson's stochastic foundation of quantum mechanics. In this context Derakhshani \cite{Derakhshani2015ab} has proposed a solution based on \textit{Zitterbewegung} at the fundamental level.

It has been tacitly assumed here and in the various works mentioned above that without the Wallstrom condition such a wave function-free formulation cannot account for the empirical data obtained from experiments involving microscopic systems. This assumption may ultimately prove accurate. However, it warrants investigation. Below we will investigate the consequences of relaxing the Wallstrom condition. In preparation to do so, we will first make some formal observations.

If a function $S$ is defined on the configuration space $\mathcal{C}$, that is, $S:\mathcal{C}\rightarrow\mathbb{R}$, and $m\vec{v}=\nabla S$, then for \textit{any} loop $\mathcal{L}$ in $\mathcal{C}$ it is the case that 
\begin{equation} \oint_{\mathcal{L}} m\vec{v}\cdot\vec{d\ell} = 0.\end{equation}
Therefore, $S$ cannot be defined globally on $\mathcal{C}$ if the integral is to match the corresponding integral obtained with the gradient of the phase of the wave function. This result is not spectacular but it does imply that $S$ cannot be thought of as a classical Hamilton Principal function.

However, $S$ may be defined \textit{locally} in $\mathcal{C}$, that is, for some open, simply connected neighbourhood $\mathcal{D}\subset\mathcal{C}$, provided $\mathcal{D}$ does not contain a point where the density $\rho$ vanishes. Indeed, in the standard quantum description of $N$ particles the phase of the wave function $\psi$ is continuous everywhere except at nodes (points where $\psi=0$), at which it is undefined. In quantum mechanics, analytic extension of the phase through the node is only possible if the integral along a loop that encircles the node vanishes. In the absence of a wave function we will use the term ``node'' to refer to a point where $\rho=0$.

The integral is invariant under continuous deformations of $\mathcal{L}$ provided the deformation does not move $\mathcal{L}$ across a node. Any loop that does not encircle a node could therefore be shrunk to a point and the corresponding integral must therefore vanish. ``Encircling'' does, of course, only make sense for nodes that are $(n-2)$-dimensional (points in two-dimensional space, lines in three-dimensional space, etc.) --- nodes of lower dimension would allow for continuous deformation of $\mathcal{L}$ to a point (such as point node in three-dimensional space). However, $n-2$-dimensional nodes occur generically in quantum mechanics since $\psi=0$ constitutes two real conditions. In density-based approaches, $\rho$ vanishes and is a minimum (assuming differentiability of $\rho$), which also constitutes two conditions. Hence an $(n-2)$-dimensional nodes occur generically here also. Spherical harmonics (e.g.\ in the context of hydrogen-like atoms) present a textbook example of this: States where the quantum number usually denoted by $m$ is non-zero have a one-dimensional node along one axis, while states with $m=0$ do not, and the loop integral of $\nabla S$ is proportional to $2\pi\hbar\cdot m$. Edge and screw dislocations (see \cite{Holland1993}, sec.\ 4.11) present another.

In the numerical investigations below we therefore exclusively focus on scenarios with nodes.

One further remark: In quantum mechanics, in any closed system the value of the loop integral eq.\ (\ref{eq.WallstromCondition}) of $\vec{v}$ around a node cannot change over time. We refer to $n$ in eq.\ (\ref{eq.WallstromCondition}) as the \textit{vorticity} of the node. Change is impossible because kinematically possible values are discrete but evolution of $\psi$ is continuous. It is therefore impossible for a system to move from one vorticity to another. Some useful quantum models allow discrete jumps between states of different vorticity but these constitute effective models of some underlying theory, not a fundamental description (e.g.\ a Hamiltonian used to calculate transition probabilities between orbitals). This result no longer holds in the CSD formulation. There is no \textit{kinematical} reason the vorticity of a system cannot change. Whether or not it \textit{does} change becomes a matter of the \emph{dynamics}. Further, in $\psi$-based formulations nodes cannot be created or destroyed except via pair creation or annihilation of nodes of opposite vorticity \cite{Underwood2018a}.


\section{Analytical and numerical investigation of non-quantum states}\label{sec.numericalanalysis}

Relaxing the Wallstrom condition opens up a broader physics. Broadly speaking, there are two paths forward in order to investigate the dynamics of this physics. First, one can directly consider the density version of the quantum Hamilton-Jacobi equation (or alternatively Newton's Law with the added quantum potential) in conjunction with the continuity equation for $\rho$. This system of equations determining $\rho$ and the vector field $\vec{v}$ on configuration space is non-linear and general properties are difficult to determine analytically. This approach does lend itself to numerical investigation however. Second, one can re-introduce a "wave function-like" function that is multi-valued on the configuration space. Nodes are analogues of branch points (or higher-dimensional equivalents) of this function. This line of inquiry will be explored further in future investigations.

For this paper we report on the numerical investigation into the Wallstrom-relaxed CSD formulation based on the first approach. Numerical investigation involves choosing an initial state and fixed grid size then numerically calculating initial values for $Q$, $\rho = |\psi|^2$ and $v =  \frac{\nabla S}{m}$ for each location on the grid. Evolution follows equations (1) and (2), and trajectory tracking is done with a 4th order Runge-Kutta method (see appendix \ref{sec.appendix}).

One approach we pursue here is to begin with a quantum state and add a small non-quantum perturbation. We then track how the size of this perturbation evolves over time. In order to track the degree of ``non-quantumness'', that is, the deviation from integer vorticity, we introduce a parameter $\nu$ as
\begin{equation} \nu = \frac{1}{2\pi\hbar}\left(\oint_0^{2\pi} m\vec{v}\cdot \vec{d\ell}\right)-n,\end{equation}
where $n$ is the (integer) vorticity corresponding to the value of the loop integral of the initial quantum state relative to which the non-quantumness is measured. For example, in two-dimensional scenarios we begin with initial fields $\rho$ and $\vec{v}$ that correspond to a quantum state --- field values that could have been derived from a single-valued wave function --- and then add a small perturbation to the velocity field of the form
\begin{equation} \vec{v}_P = \hbar\nu_0\frac{\hat\theta}{r} = \frac{\hbar\nu_0}{x^2+y^2}\twovector{-y}{x}, \label{nonQM-perturbation}\end{equation}
where the coordinates are assumed centered on the node. Note that this velocity-field perturbation may be expressed as the gradient of a function in any open simply-connected neighborhood that does not encompass the origin (the node), namely as
\begin{equation} \vec{v}_P = \nabla S_P,\qquad\text{where } S_P=\hbar\nu_0\,\arctan\left(\frac{y}{x}\right).\end{equation}
As expected, the function $S_P$ would be multi-valued if analytically continued over the entire configuration space minus the node. The parameter $\nu_0$ is the initial degree of non-quantumness, $\nu_0 = \nu(0)$. Given that any velocity perturbation we consider is curl-free except at the node, this velocity-field perturbation is generic: Any other perturbation could be expressed as the gradient of the sum of $S_P$ and a function that is analytic on the entire configuration space and therefore leaves the vorticity unchanged.


\section{Case studies: The 2D Harmonic Oscillator} \label{sec.casestudies}

The harmonic oscillator presents a paradigm example when studying both classical and quantum systems. Its relevance reaches from small oscillations of mechanical systems to quantum field theory and cosmological perturbations. In the context of the present investigation the one-dimensional harmonic oscillator is of little interest since in one dimension nodes cannot be encircled by loops. However, the two-dimensional harmonic oscillator is sufficient to display non-quantum behavior. We study it in this section. The algorithmic details are briefly described in appendix \ref{sec.appendix}.

In standard quantum mechanics, the one-dimensional oscillator has eigenstates
\begin{equation*} \psi_j(x) = \left(\frac{m\omega}{\pi \hbar}\right)^{\frac{1}{4}} \frac{1}{\sqrt{2^j j!}}H_j\left(\sqrt{\frac{m\omega}{\hbar}}x\right)e^{-\frac{m\omega}{2\hbar}x^2} \end{equation*}
for $j=0,~1,~2,~\dots$. A two-dimensional basis for the space of all allowed states is formed by
\begin{equation} \psi_{jk} = \psi_j(x)\psi_k(y). \end{equation}
We are interested in states with nodes, since such states allow generalization to the non-quantum realm. A node here is an $(n-2)=0$-dimensional locus, a point. In particular, we require nodes of non-zero vorticity. Such states may be constructed as the sum of two or more basis states with relative phase factors. For example,
\begin{equation} \psi = \psi_{01} + i\psi_{10}. \label{SHOproductstate}\end{equation}
It is straightforward to verify by explicit calculation that the vorticity $n$ of such a state is a non-zero integer (here $n=1$). Computed trajectories orbit the node. Speed falls off with distance from the node. Testing a variety of such states that fall within configurations accessible by standard quantum mechanics served to corroborate reliability of the numerical computation. Explicit computation in CSD (via $\rho$ and $\vec{v}$ as fundamental quantities rather than $\psi$) shows $n$ remains constant, as one would expect given that the initial state of the system has a quantum-mechanical counterpart. That $n$ is constant is however a result of the \textit{dynamics} in CSD, unlike in standard quantum mechanics, where non-integer $n$ is kinematically impossible.

To these states we added a perturbation of the form of equation (\ref{nonQM-perturbation}). The stability of the vorticity remained unchanged. Note though that a product state of the general form (\ref{SHOproductstate}) does not remain a stationary state if the velocity field is perturbed while the density function is not.\footnote{See the discussion on stationary states in section \ref{sec.stationarystates}.} In particular, an increase in the initial speed must lead to an initial outwards motion of trajectories from the node (since the instantaneous acceleration remains unchanged upon this perturbation) and hence a change in the radial dependence of $\rho(r)$.

The stability of the vorticity extends to situations of time-dependent potentials. We investigated several choices of $V(x,t)$ to simulate the influence of an external interaction, such as the temporary proximity of another particle. In all cases the vorticity remained unaffected. We also added non-quantum perturbations to other scenarios in which nodes occur, such as the `dislocated wave' (see \cite{Holland1993}, 4.11) and an infinite circular well, and found similar results.

The result of the simulation is the demonstration of the \textit{dynamical} stability of the degree of non-quantumness of this simple system, including under various forms of perturbations that can be taken to simulate interactions.


\section{Stationary States in CSD Frameworks}\label{sec.stationarystates}

Stationary states in quantum mechanics correspond to the energy eigenfunctions of the system, the set of which is discrete in scenarios such as this one. An analysis of a textbook derivation of the 2D harmonic oscillator, the circular infinite well, or the hydrogen atom, for example, shows that the discreteness arises exactly as a result of imposing the Wallstrom condition, or, in other words (in standard quantum terminology), as a result of the single-valuedness of the wave function. Every quantum stationary state is also stationary in CSD (in the sense that $\rho(x)$ and $\vec{v}(x)$ are time independent), but the absence of the Wallstrom condition creates a continuous rather than discrete infinity of such basis states, characterized by continuous values of vorticity $n$. There is no wave function in the quantum description of particular system that corresponds to stationary states of non-integer $n$. 

It is conceivable however to develop a pseudo-quantum description of such states in terms of multi-valued wave functions (specifically, multi-valued phases), or states defined on configuration spaces with a branch cut, with the node as its branch point. ``Point'' here is to be understood as defined on a suitably chosen two-dimensional subspace. This we leave to future work.

The analysis of non-integer vorticity states presents a mathematical challenge since the radial functions are, depending on the particular details of the scenario, various special functions of non-integer order. In the harmonic oscillator studied in section \ref{sec.casestudies} these functions are the Hermite polynomials. In other cases the properties of non-integer-order Bessel functions of the first kind (infinite well) or non-integer-order generalized Laguerre polynomials (hydrogen atom) need to be studied in more depth.

In most of the literature on special functions non-integer orders are delegated to a footnote or very brief discussion only.\footnote{Arfken (\cite{Arfken1985}), for example, only dedicates two paragraphs to the subject on non-integer Bessel functions over the course of sixty pages on the subject as a whole.} Series or integral representations for such functions exist that are well-defined for non-integer cases, often in terms of the Gamma function (e.g.\ \cite{Arfken1985}). For Bessel functions of the first kind, for example (relevant to the circular well),
\begin{equation} \label{eq.BesselSeries}
    J_n(x) = \sum_{s=0}^\infty \frac{(-1)^s}{s!\Gamma(n+s+1)}\left(\frac{x}{2}\right)^{n+2s}
\end{equation}
holds for non-integer $n$. However, it is interesting to note that there are \textit{qualitative} differences between non-integer and integer-order special functions. For the functions \ref{eq.BesselSeries} only integer-order ones are entire, while non-integer ones are multi-valued with a singularity at zero when defined over $\mathbb{C}$. 

We posit here, speculatively, the possibility that this line of investigation may yet reveal insights into what makes integer vorticity special, and perhaps ultimately explain the empirical absence of non-integer vorticity even in CSD frameworks. The thought here is that integer vorticity arises because when expanded around the node the \textit{radial} functions $\rho(r)$, which are linked via the dynamics to the velocity field, have special properties in exactly those cases in which the integral \ref{eq.WallstromCondition} has integer $n$. A more detailed analysis of the stationary states in CSD frameworks is required. This constitutes future work.


\section{Conclusion}\label{sec.conclusion}

In all cases we studied the degree of non-quantumness is conserved. This is akin to angular momentum conservation around the node across the ensemble. The conservation might therefore not come as a surprise, although it must not be presupposed given that the theory studied here is notably distinct from quantum mechanics. We noted that the \textit{kinematic} conservation of vorticity is not a feature of CSD theories. Here angular momentum \textit{across the ensemble} is dynamically conserved. From the CSD perspective, it can be argued that this result indicates that a kinematical constraint of the vorticity is unnecessary structure. However, this argument is undermined by the fact that the absence of non-integer vorticity in experimental data is yet to explained. Barring the attempts outlined in section \ref{sec.wallstromnonquamtum} or section \ref{sec.stationarystates} such a justification remains outstanding.

In standard quantum mechanics the experimental evidence for quantized vorticity is the outcomes of what one would term ``angular momentum measurements'' --- really quantized eigenvalues of a mathematical operator one calls the ``angular momentum operator.'' This interpretation of mathematical operators as physical measurements makes (even) less sense in CSD frameworks, but the interpretation and terminology is irrelevant for what is observed and what standard quantum mechanics does, and what CSD frameworks as they stand right now do not predict.

Of course, more complex and higher-dimensional systems are yet to be investigated, so the generalization of the results of the numerical investigation is conjecture only. However, the insights so far stifle hope that the Wallstrom challenge might have been resolved dynamically with the quantized vorticities of quantum mechanics (and pilot-wave theory) as some form of stable equilibrium towards which systems tend generically under their own evolution. Any proposal for a CSD-type theory must therefore explicitly address the Wallstrom condition.

Investigating CSD theories in the non-quantum regime is, of course, not limited to merely tracking the degree of non-quantumness. They may have interesting dynamics in their own right with qualitative differences to dynamics of the quantum regime. A study of non-quantum stationary states as outlined in section \ref{sec.stationarystates} may provide a systematic entry point towards further study of this nature. 

\appendix
\section{Computational Methods} \label{sec.appendix}
The data and code used in this paper are available at \href{https://github.com/mscoggs/no_wave_qm}{https://github.com/mscoggs/no\_wave\_qm}. Simulation is performed on a uniform square grid, where initial density and velocity field are specified, then $Q$, $\rho$, $v$ and $V(x_i)$ are updated at each time step. In order to ensure computational reliability of the result each simulation is repeated with smaller time step sizes until a further decrease in the time step has negligible effect on the evolution. The algorithm also allows tracking of individual trajectories based on user-selected initial positions. Trajectory evolution is calculated using a fourth-order Runge-Kutta method.


\setlength{\bibsep}{0pt plus 0.3ex}
\small
\bibliographystyle{unsrtnat}	
\bibliography{NonQuantum-Bibloi}	

\begin{thebibliography}{30}
\providecommand{\natexlab}[1]{#1}
\providecommand{\url}[1]{\texttt{#1}}
\expandafter\ifx\csname urlstyle\endcsname\relax
  \providecommand{\doi}[1]{doi: #1}\else
  \providecommand{\doi}{doi: \begingroup \urlstyle{rm}\Url}\fi

\bibitem[Wallace(2008)]{Wallace2008}
D.~Wallace.
\newblock {Philosophy of Quantum Mechanics}.
\newblock In \emph{{The Ashgate Companion to Contemporary Philosophy of
  Physics}}. Routledge (London), 2008.

\bibitem[Schlosshauer(2011)]{Schlosshauer2011}
M.~Schlosshauer.
\newblock \emph{{Elegance and Enigma}}.
\newblock Springer, 2011.

\bibitem[Tipler(2010)]{Tipler2010}
F.~Tipler.
\newblock {Hamilton-Jacobi many-worlds theory and the Heisenberg uncertainty
  principle}.
\newblock 2010.
\newblock quant-ph: 1007.4566.

\bibitem[Sebens(2014)]{Sebens2014}
C.~Sebens.
\newblock {Quantum Mechanics as Classical Physics}.
\newblock \emph{quant-ph}, 2014.

\bibitem[Hall et~al.(2014)Hall, Deckert, and Wiseman]{HallDeckertWiseman2014}
M.~Hall, D.-A. Deckert, and H.~Wiseman.
\newblock {Quantum phenomena modelled by interactions between many classical
  worlds}.
\newblock \emph{Phys. Rev. X}, 4:\penalty0 041013, 2014.
\newblock quant-ph: 1402.6144.

\bibitem[Schmelzer(2011{\natexlab{a}})]{Schmelzer2011a}
I.~Schmelzer.
\newblock {The paleoclassical Interpretation of quantum theory}.
\newblock \emph{quant-ph: 1103.3506}, 2011{\natexlab{a}}.

\bibitem[Smolin(2015)]{Smolin2015}
L.\ Smolin.
\newblock {Quantum mechanics and the principle of maximal variety}.
\newblock 2015.
\newblock quant-ph: 1506.02938.

\bibitem[Bostr{\"o}m(2015)]{Bostrom2014}
K.~J.\ Bostr{\"o}m.
\newblock {Quantum mechanics as a deterministic theory of a continuum of
  worlds}.
\newblock \emph{Quantum Studies: Mathematics and Foundations}, 2, 2015.
\newblock quant-ph: 1410.5653.

\bibitem[Roser(2015)]{Roser2015_TrajGeometry}
P.~Roser.
\newblock {Quantum mechanics as the dynamical geometry of trajectories}.
\newblock 2015.
\newblock \href{https://arxiv.org/abs/1507.08975}{quant-ph: 1507.08975}.

\bibitem[Pusey et~al.(2012)Pusey, Barrett, and
  Rudolph]{PuseyBarrettRudolph2012}
M.~Pusey, J.~Barrett, and T.~Rudolph.
\newblock {On the reality of the quantum state}.
\newblock \emph{Nature Physics}, 8:\penalty0 475, 2012.
\newblock quant-ph: 1111.3328.

\bibitem[Everett(1956)]{Everett1956}
H.~Everett.
\newblock \emph{{The theory of the universal wavefunction}}.
\newblock PhD thesis, Princeton University, 1956.

\bibitem[Madelung(1927)]{Madelung1926}
E.~Madelung.
\newblock {Quantentheorie in hydrodynamischer Form}.
\newblock \emph{Zeitschrift f\"ur Physik}, 40:\penalty0 322--326, 1927.

\bibitem[Bohm(1952{\natexlab{a}})]{Bohm1952a}
D.~Bohm.
\newblock {A Suggested Interpretation of the Quantum Theory in Terms of
  `Hidden' Variables. I}.
\newblock \emph{Physical Review}, 85:\penalty0 166--179, 1952{\natexlab{a}}.

\bibitem[Bohm(1952{\natexlab{b}})]{Bohm1952b}
D.~Bohm.
\newblock {A Suggested Interpretation of the Quantum Theory in Terms of
  `Hidden' Variables. II}.
\newblock \emph{Physical Review}, 85:\penalty0 180--194, 1952{\natexlab{b}}.

\bibitem[Holland(1993)]{Holland1993}
P.~R. Holland.
\newblock \emph{{The Quantum Theory of Motion: an Account of the
  de~Broglie-Bohm Causal Interpretation of Quantum Mechanics}}.
\newblock Cambridge University Press, 1993.

\bibitem[D\"urr et~al.(1992)D\"urr, Goldstein, and
  Zangh\'i]{DuerrGoldsteinZanghi1992}
D.~D\"urr, S.~Goldstein, and N.~Zangh\'i.
\newblock {Quantum Equilibrium and the Origin of Absolute Uncertainty}.
\newblock \emph{Journal of Statistical Physics}, 67:\penalty0 843--907, 1992.

\bibitem[Valentini(1991)]{Valentini1991a}
A.~Valentini.
\newblock {Signal locality, uncertainty, and the subquantum H-theorem. I}.
\newblock \emph{Physics Letters A}, 156:\penalty0 5, 1991.

\bibitem[Valentini and Westman(2004)]{ValentiniWestman2004}
A.~Valentini and H.~Westman.
\newblock {Dynamical origin of quantum probabilities}.
\newblock \emph{Proceedings of the Royal Society A}, 461:\penalty0 253--272,
  2004.

\bibitem[Colin and Valentini(2014)]{ColinValentini2014}
S.~Colin and A.~Valentini.
\newblock {Instability of quantum equilibrium in Bohm's dynamics}.
\newblock \emph{Proceedings of the Royal Society A}, 470:\penalty0 20140288,
  2014.
\newblock quant-ph: 1306.1576.

\bibitem[Poirier(2010)]{Poirier2010}
B.~Poirier.
\newblock {Bohmian Mechanics without pilot waves}.
\newblock \emph{Chemical Physics}, 370:\penalty0 4, 2010.

\bibitem[Wya()]{Wyatt2005}
\emph{Quantum Dynamics With Trajectories}.

\bibitem[Dirac(1931)]{Dirac1931}
P.A.M. Dirac.
\newblock {Quantized Singularities in the Electromagnetic Field}.
\newblock \emph{Proceedings of the Royal Society A}, page~60, 1931.

\bibitem[Wallstrom(1994)]{Wallstrom1994}
T.~C. Wallstrom.
\newblock {Inequivalence between the Schr\"odinger equation and the Madelung
  hydrodynamic equations}.
\newblock \emph{Physical Review A}, 49:\penalty0 1613, 1994.

\bibitem[Schmelzer(2011{\natexlab{b}})]{Schmelzer2011b}
I.~Schmelzer.
\newblock {A solution for the Wallstrom problem of Nelsonian stochastics}.
\newblock \emph{quant-ph: 1101.5774}, 2011{\natexlab{b}}.

\bibitem[Santamato(1984{\natexlab{a}})]{Santamato1984}
E.~Santamato.
\newblock Geometric derivation of the schr{\"o}dinger equation from classical
  mechanics in curved weyl-spaces.
\newblock \emph{Physical Review D}, 29\penalty0 (2):\penalty0 216, January
  1984{\natexlab{a}}.

\bibitem[Santamato(1984{\natexlab{b}})]{Santamato1984b}
E.~Santamato.
\newblock {Statistical interpretation of the Klein-Gordon equation int erms of
  the spacetime Weyl curvature}.
\newblock \emph{Journal of Mathematical Physics}, 25:\penalty0 2477,
  1984{\natexlab{b}}.

\bibitem[Goldstein et~al.(2000)Goldstein, Poole, and Safko]{Goldstein3rdEd}
H.~Goldstein, C.~Poole, and J.~Safko.
\newblock \emph{{Classical Mechanics, 3rd ed.}}
\newblock Addison-Wesley, 2000.

\bibitem[Derakhshani(2015, 2016)]{Derakhshani2015ab}
M.~Derakhshani.
\newblock {A Suggested Answer To Wallstrom's Criticism: Zitterbewegung
  Stochastic Mechanics I + II}.
\newblock \emph{quant-ph: 1510.06391, 1607.08838}, 2015, 2016.

\bibitem[Underwood(2018)]{Underwood2018a}
N.~Underwood.
\newblock {Extreme quantum nonequilibrium, nodes, vorticity, drift, and
  relaxation retarding states}.
\newblock \emph{Journal of Physics A}, 51:\penalty0 055301, 2018.
\newblock quant-ph: 1705.06757.

\bibitem[Arfken(1985)]{Arfken1985}
G.~Arfken.
\newblock \emph{{Mathematical Methods for Physicists, 3rd ed.}}
\newblock Academic Press, 1985.

\end{thebibliography}
\end{multicols}

\end{document}